\documentclass[pdflatex,sn-mathphys-num]{sn-jnl}

\usepackage{graphicx}
\usepackage{amsmath,amssymb,amsfonts}
\usepackage{amsthm}
\usepackage{mathrsfs}
\usepackage[title]{appendix}
\usepackage{xcolor}
\usepackage{textcomp}
\usepackage{booktabs}
\usepackage{listings}
\usepackage[T1]{fontenc}
\usepackage[utf8]{inputenc}
\usepackage{lmodern}
\usepackage{verbatim}

\begin{document}

\title[Article Title]{From Lasers to Photon Bose--Einstein Condensates: A Unified Description via an Open-Dissipative Bose--Einstein Distribution}

\author*[1]{\fnm{Joshua} \sur{Krauß}}\email{jkrauss@rptu.de}

\author[1]{\fnm{Enrico} \sur{Stein}}\email{steinenrico94@gmail.com}

\author[1]{\fnm{Axel} \sur{Pelster}}\email{axel.pelster@rptu.de}

\affil*[1]{\orgdiv{Department of Physics and Research Center OPTIMAS}, \orgname{University Kaiserslautern-Landau}, \orgaddress{\street{Erwin-Schrödinger Straße 46}, \city{Kaiserslautern}, \postcode{67663}, \state{Rhineland-Palatinate}, \country{Germany}}}

\abstract{Photon condensation was first experimentally realized in 2010 within a dye-filled microcavity at room temperature. 
Since then, interest in the field has increased significantly, as a photon Bose-Einstein condensate (BEC) represents a prototypical driven-dissipative system. 
Here, we investigate how its inherent open nature influences the condensation process both quantitatively and qualitatively. 
To this end, we consider a mean-field model, which can be derived microscopically from a Lindblad master equation. 
The underlying rate equations depend on various external parameters such as emission and absorption rates of the dye molecules as well as the cavity photon loss rate. 
In steady state, we obtain an open-dissipative Bose-Einstein distribution for the mode occupations.
The chemical potential of this distribution depends on the occupations of the dye molecules in both their ground and excited state and must therefore be determined self-consistently. 
We find that the resulting photon distribution is strongly influenced by the driven-dissipative parameters. 
Based on this result, we identify the main differences between a photonic BEC, an atomic BEC, and a laser.}
\maketitle
\newpage
\section{Introduction}\label{sec1}
The macroscopic quantum phenomenon Bose--Einstein condensation (BEC) was theoretically predicted 
by A. Einstein~\cite{Einstein_Berichte1924} in 1925 based on a quantum statistical derivation of Planck's black-body radiation formula by S.N. Bose~\cite{Bose_Zeitschrift1924} in 1924. 
The first experimental realization was achieved in 1995 in dilute atomic vapors of rubidium and sodium at ultracold temperatures in the nanokelvin range~\cite{Cornell_Science1995,Ketterle_PRL1995}. 
Since then, atomic BECs have been intensively studied. 
Today, Bose--Einstein condensation is a well-understood phenomenon, which is frequently employed in a wide range of experimental platforms~\cite{Pethick_Smith,Pitaevskii_Stringari}.\\
Photonic systems are of particular interest because of their open and dissipative character. 
These platforms have attracted considerable attention over the past decades, both from the point of view of fundamental research and practical applications. 
A prominent example is the laser~\cite{Haken_book1981,Haken_book1983,Haken_book1985,Meschede_book2017}, which has played a central role in developing our understanding of non-equilibrium phase transitions and has found widespread use in industrial applications. 
Another promising research direction is the photon BEC. 
It emerges from an equilibrium phase transition and was first realized in a dye-filled microcavity at room temperature~\cite{Weitz_Nature2010,Nyman_PRA2015,Oosten_PRA2018} and, more recently, also in a vertical cavity surface emitting laser (VCSEL)~\cite{Fainstein_2024,Schofield_NaturePhoton2024,Pieczarka_NaturePhoton2024}.
In this work, we focus on the case of the dye-filled microcavity.\\ 
Although both lasers and photon BECs involve ensembles of photons, they exhibit fundamental differences. In the simplest approximation, a laser consists of an ensemble of two-level atoms coupled to a single light field. 
The atoms are externally pumped and can decay into non-cavity modes, while the light field undergoes cavity losses. 
As derived by H. Haken~\cite{Haken_book1981,Haken_book1983,Haken_book1985}, the most basic description of such a system employs rate equations, which capture the dynamics of the ground and excited atomic populations together with the photon field. 
These rate equations can be derived microscopically using a master-equation approach~\cite{Haken_book1985,Cummings_IEEE1963}.\\
A photon BEC is conceptually similar but differs in key aspects. 
It consists of a curved-mirror cavity, which modifies the photon dispersion from linear to quadratic and introduces a finite ground-state energy. 
The mirrors also provide a harmonic trapping potential for the cavity photons. 
The cavity is filled with a dye that is incoherently pumped. 
This enables effective photon-photon interaction via Kerr and thermo-optic effects~\cite{Boyd_book2008}. More importantly, repeated absorption and re-emission processes allow the dye to act as both a heat and a particle reservoir, leading to a thermalization of the photon gas at room temperature~\cite{Weitz_Nature2010}. 
This thermalization mechanism, combined with the use of high-quality mirrors~\cite{Weitz_Nature2010,Sazhin_Natcom2024}, fundamentally distinguishes a photon BEC from a laser. 
Nevertheless, as microscopically derived with a master-equation approach, it was shown in  Refs.~\cite{Kirton_PRL2013,Kirton_PRA2015,Radonjic_NJP2018,Erglis_PRL2024} that the resulting dynamics can be effectively described by rate equations, which show good agreement with experimental observations~\cite{Ozturk_PRA2019,Sazhin_Natcom2024,Erglis_PRL2025}.\\
In steady state, it was demonstrated by P. Kirton and J. Keeling in~\cite{Kirton_PRL2013} that the photon BEC follows a generalized BEC distribution. 
This distribution extends the standard equilibrium form by including non-equilibrium parameters and reduces to the standard Bose--Einstein distribution in the equilibrium limit. 
Thus, strictly speaking, photon thermalization does not constitute an equilibrium phase transition. 
Despite its non-equilibrium character, the photon BEC has, in all previous applications, been sufficiently approximated as an ideal, non-interacting gas undergoing an equilibrium phase transition~\cite{Klaers_Nature2010,Klaers_APB2011,Klaers_PRL2012,Schmitt_PRA2015,Damm2016,Busley_PRA2023}. 
Hence, this raises the fundamental question of how the non-equilibrium nature of the photon BEC affects the validity of the equilibrium approximation.\\
To address this question, we analyze the rate-equation description of the photon BEC and investigate the role of non-equilibrium corrections. 
To this end, Section~\ref{sec2} starts with introducing the underlying mean-field model, the steady state of which is given by an open-dissipative BEC distribution. 
Additionally, we examine the single-mode limit, which contains the special case of a laser and qualitatively coincides with the results derived by H. Haken in Ref.~\cite{Haken_book1985}. 
Since no exact analytical solution of the general model is available, we proceed with numerical simulations. 
In view of this, Section~\ref{sec3} summarizes the relevant parameters, which are later used in the simulations. 
Section~\ref{sec4} presents the numerical results for three representative cases. 
First, we show that including a sufficiently large number of photon modes allows the system to approach the thermodynamic limit. 
Second, we investigate how the non-equilibrium parameter in the open-dissipative distribution influences the system by varying its degree of openness. 
Third, we examine the effect of different dye solutions. 
Finally, Section~\ref{sec5} summarizes our findings and provides an outlook.
\section{Rate Equation Description}\label{sec2}
We start by introducing the rate equation model for describing the thermalization of a photon Bose-Einstein condensate. 
A subsequent steady-state analysis yields a generalized Bose-Einstein distribution that accounts for the open-dissipative character of the system. 
In particular, we focus on discussing how the corresponding equation of state depends on the respective open-dissipative parameters. 
Finally, we specialize the rate equation model to the case of a single mode, determine its exact analytical solution, and discuss the connection to a laser.
\subsection{Model}\label{subsec21}
One of the simplest, yet most profound, ways to describe the thermalization process of a photon BEC is through rate equations. 
These can be derived microscopically~\cite{Radonjic_NJP2018,Kirton_PRL2013,Kirton_PRA2015,Ozturk_PRA2019,Bode_PRR2024} and allow for a phenomenological interpretation based on the fundamental Einstein processes for the light-matter interaction.\\
The model distinguishes two different kinds of molecular states, namely the electronic ground and excited state, with their corresponding occupation numbers $M_1$ and $M_2$. 
Furthermore, one considers the respective photon modes with index $\ell=0,1,2,\ldots$, frequency $\omega_\ell$, and degeneracy $g_\ell$, each being occupied by $N_\ell$ photons. 
Molecules and photons interact through the fundamental Einstein processes of absorption and emission. They are characterized by the coefficients $B_{12}$ and $B_{21}$, which are connected  by a Boltzmann factor due to the Kennard--Stepanov (KS) relation~\cite{Kennard_PR1918,Stepanov_SR1957}
\begin{equation}
	\label{eq:KS_relation}
	B_{12}(\omega_\ell) = B_{21}(\omega_\ell) e^{\hbar\beta(\omega_\ell - \omega_\text{ZPL})}\,.
\end{equation}
Here, $\beta=1/k_{\rm B} T$ represents the inverse temperature and $\omega_\text{ZPL}$ denotes the zero-phonon line frequency. 
Note that Eq.~\eqref{eq:KS_relation} can also be derived microscopically within a master equation approach by coupling the dye molecules to their own vibrational thermal bath~\cite{Kirton_PRL2013,Kirton_PRA2015,Radonjic_NJP2018}.
Furthermore, the finite photon lifetime is modeled by cavity losses at rate $\Gamma_\text{c}$, which are compensated by an external laser field with pumping rate $p$. 
Furthermore, radiationless molecular transitions from the excited to the ground state provide an additional loss channel, which is quantified by the rate $\Gamma_\text{nr}$. Taking all these processes together leads to the following system of rate equations:
\begin{align}
	\label{eq:photon_rate}
	\dot{N}_\ell &= \Big[B_{21}(\omega_\ell) M_2 - B_{12}(\omega_\ell) M_1 -\Gamma_\text{c}(\omega_\ell) \Big]N_\ell + B_{21}(\omega_\ell) M_2\,,\\
	\label{eq:molecule_rate}
	\dot{M}_2 &= p M_1 -\Gamma_\text{nr} M_2 - \sum_{\ell=0}^{\infty} \left\{\Big[B_{21}(\omega_\ell) M_2 - B_{12}(\omega_\ell) M_1\Big]N_\ell + B_{21}(\omega_\ell) M_2\right\}\,,\\
	\label{eq:molecule_cons}
	M &= M_1 + M_2\,.
\end{align}
Equation~\eqref{eq:photon_rate} contains the following contributions: the first two terms describe stimulated emission and absorption of photons by the molecules, whereas the third term represents frequency-dependent cavity losses, and the last one corresponds to spontaneous emission. 
In Eq.~\eqref{eq:molecule_rate}, the first two terms arise from external pumping and radiationless decay, while the remaining terms capture the coupling of photons and molecules through absorption and emission. 
The system is completed by Eq.~\eqref{eq:molecule_cons}, which accounts for the conservation of the total molecule number $M$, and is equivalent to $\dot{M}_2 = -\dot{M}_1$. 
It should be noted that, unlike the laser rate equations, where $B_{21} = B_{12}$~\cite{Haken_book1985}, the photon BEC case requires the KS relation~\eqref{eq:KS_relation}, since the molecules act as both a heat and a particle reservoir.
\subsection{Steady-State}\label{subsec22}
We now examine the steady state solution of the rate equations~\eqref{eq:photon_rate}--\eqref{eq:molecule_cons}. 
To this end, we start with the photon mode equations~\eqref{eq:photon_rate}, which yield together with the KS relation~\eqref{eq:KS_relation} the open-dissipative Bose--Einstein distribution~\cite{Kirton_PRL2013}
\begin{equation}
	\label{eq:open_BEC}
	N_\ell = \frac{1}{e^{\beta(\hbar\omega_\ell - \mu)} - 1 + \frac{\Gamma_c(\omega_\ell)}{B_{21}(\omega_\ell)M_2}}\,,
\end{equation} 
where the chemical potential turns out to be  
\begin{equation}
	\label{eq:chemical_pot}
	\mu = \hbar\omega_\text{ZPL} - \frac{1}{\beta}\ln\left(\frac{M_1}{M_2}\right)\, .
\end{equation}
Equation~\eqref{eq:open_BEC} represents a modification of the standard Bose--Einstein distribution, adapted for photon condensates by incorporating the driven-dissipative nature of the system. 
In this expression, the additional denominator term  compares the time scale for spontaneous emission $t_\text{em}=1/B_{21}M_2$ with the photon cavity lifetime $t_\text{ph} = 1/\Gamma_\text{c}$. 
Thus, the ratio $t_{\rm em}/t_{\rm ph}= \Gamma_{\rm c}/B_{21}M_2$ is expected to be small if many photon emission–absorption cycles occur before the photons escape the cavity, and thus represents a measure of thermalization. 
Indeed, if this ratio is negligibly small, Eq.~\eqref{eq:open_BEC} reduces to the standard Bose--Einstein distribution. 
Although this condition is experimentally realized with highly reflective mirrors~\cite{Weitz_Nature2010,Sazhin_Natcom2024}, the precise influence of this additional term has never been theoretically investigated. 
Furthermore, Eq.~\eqref{eq:chemical_pot}
explicitly links the chemical potential $\mu$ to the molecule occupation numbers $M_1, M_2$, which adjust at equilibrium due to the dye-photon interaction. 
Solving the remaining equations~\eqref{eq:molecule_rate} and~\eqref{eq:molecule_cons} in steady state leads to
\begin{eqnarray}
	\label{eq:molecule_solution1}
		M_1 &=& \frac{\Gamma_\text{nr}M + \sum_{\ell=0}^{\infty}\Gamma_\text{c}(\omega_\ell)N_\ell}{\Gamma_\text{nr} + p}\,,\\
    \label{eq:molecule_solution2}
		M_2 &=& \frac{pM - \sum_{\ell=0}^{\infty}\Gamma_\text{c}(\omega_\ell)N_\ell}{\Gamma_\text{nr} + p}\, ,
\end{eqnarray}
which reveals how the chemical potential~\eqref{eq:chemical_pot} depends on the respective photon numbers $N_\ell$:
\begin{equation}
	\label{eq:molecule_frac}
\mu = \hbar \omega_{\rm ZPL} - \frac{1}{\beta} \ln\left( 
 \frac{\Gamma_\text{nr}M + \sum_{\ell=0}^{\infty}\Gamma_\text{c}(\omega_\ell)N_\ell}{pM - \sum_{\ell=0}^{\infty}\Gamma_\text{c}(\omega_\ell)N_\ell}\right)\,.
\end{equation}
Inserting the photon numbers $N_{\ell}$ from Eq.~\eqref{eq:open_BEC} into \eqref{eq:molecule_frac} reveals that the chemical potential $\mu$ follows from a self-consistency equation. 
Once $\mu$ is determined, inserting it into Eq.~\eqref{eq:open_BEC} leads to the respective photon numbers $N_{\ell}$.
\subsection{Single-Mode Limit: Photon BEC versus Laser}\label{subsec23}
Here we focus on examining the steady state of the rate equations for such a large pumping, i.e.,~such a large photon number, that practically only a single photon mode with frequency $\omega_\ell$ is coupled to the molecules. 
In this case, thermal photons do not occur, so all photons contribute to the photon Bose--Einstein condensate.
By specializing both the open-dissipative Bose--Einstein distribution~\eqref{eq:open_BEC} and the self-consistency equation~\eqref{eq:molecule_frac} for the chemical potential to a single cavity mode one obtains formally two solutions for the corresponding photon number:
\begin{eqnarray}
\nonumber	
		N_\mp &= &\frac{1}{2\left(1+\frac{B_{12}}{B_{21}}\right)}\Biggl[\frac{pM-\Gamma_{\rm c} - M\Gamma_{\rm nr}\frac{B_{12}}{B_{21}}}{\Gamma_{\rm c}} - \frac{p + \Gamma_{\rm nr}}{B_{21}} \\
		&& \mp \sqrt{\frac{4pM}{\Gamma_{\rm c}}\left(1+\frac{B_{12}}{B_{21}}\right) + \left(\frac{p + \Gamma_{\rm nr}}{B_{21}} - \frac{pM-\Gamma_{\rm c} - M\Gamma_{\rm nr}\frac{B_{12}}{B_{21}}}{\Gamma_{\rm c}}\right)^2}\,\,\Biggr]\,.
\label{eq:photon_single_mode}
\end{eqnarray}
As the solution $N_{-}$ in Eq.~\eqref{eq:photon_single_mode} becomes negative for small pumping, it is non-physical and will therefore be discarded in the following. 
In contrast to that, the solution $N_{+}$ in Eq.~\eqref{eq:photon_single_mode} depends linearly on the pumping and tends to zero for vanishing pumping, which is physically expected. 
In the opposite limit of large pumping the photon number in the condensate turns out to also depend linearly on the pumping
\begin{eqnarray}
	 \label{eq:Nminus_largep}
	N_+ &= &\frac{M - \frac{\Gamma_{\rm c}}{B_{21}} + \left| M - \frac{\Gamma_{\rm c}}{B_{21}}\right|}{2\Gamma_{\rm c}\left( 1 + \frac{B_{12}}{B_{21}} \right)} \cdot p - \frac{M\Gamma_{\rm nr}\frac{B_{12}}{B_{21}} + \Gamma_{\rm c}\left(1 + \frac{\Gamma_{\rm nr}}{B_{21}}\right)}{2\Gamma_{\rm c}\left( 1 + \frac{B_{12}}{B_{21}} \right)}\\
    &&+ \frac{\Gamma_{\rm c}\left(2M\frac{B_{12}}{B_{21}} + M + \frac{\Gamma_{\rm c}}{B_{21}}\right) + \Gamma_{\rm nr}\left(M\frac{B_{12}}{B_{21}} + \frac{\Gamma_{\rm c}}{B_{21}}\right)\left(\frac{\Gamma_{\rm c}}{B_{21}} - M\right)}{2\Gamma_{\rm c}\left( 1 + \frac{B_{12}}{B_{21}} \right)\left| M - \frac{\Gamma_{\rm c}}{B_{21}}\right|} + \mathcal{O}\left(\frac{1}{p}\right)\,.\nonumber
\end{eqnarray}
We note that for $M < \Gamma_{\rm c}/B_{21}$ the linear contribution vanishes, leading to an unphysical solution of the rate equations, since a linear dependence 
\begin{figure}[t]
	\centering
	\includegraphics[width=\columnwidth]{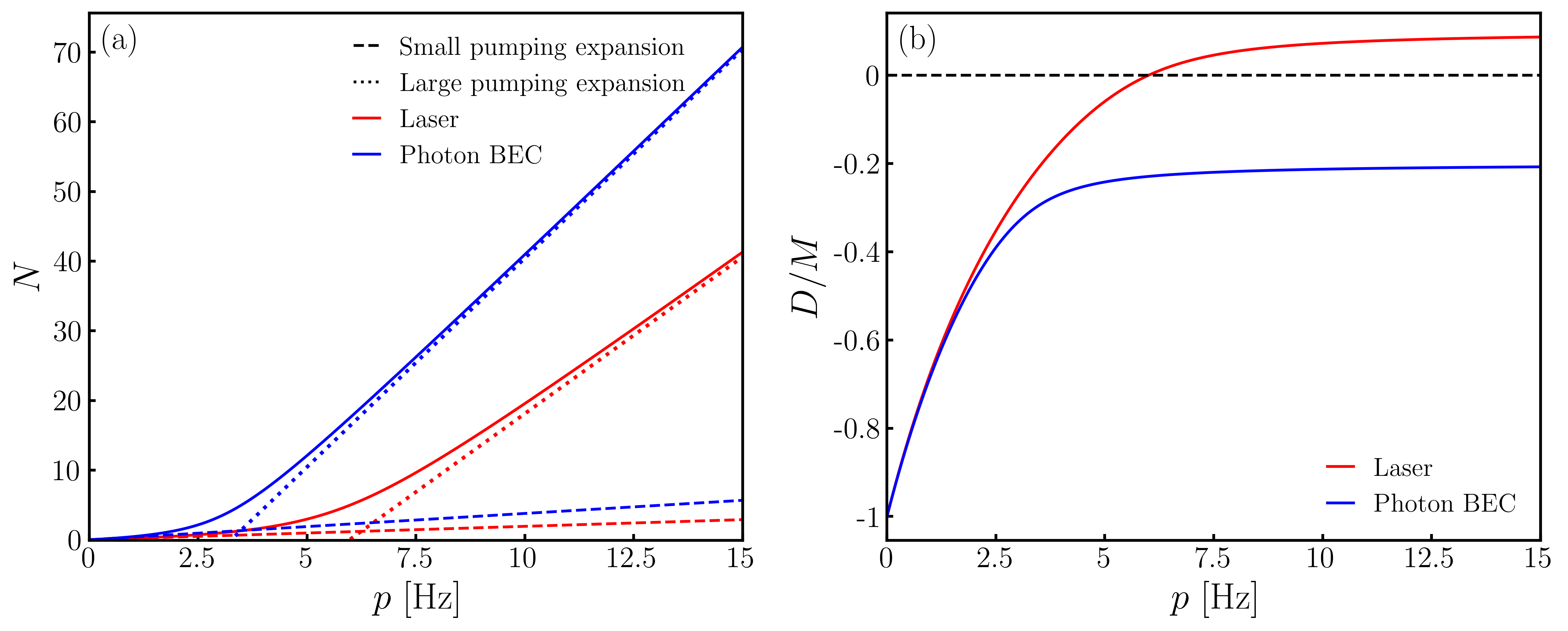}
	\caption{Panel (a) shows for the single-mode model the photon number $N$ as a function of the pumping strength $p$. Dotted  and dashed lines represent small and large pumping limits, respectively. Blue curves correspond to the photon BEC case of unequal absorption and emission rates  $B_{12} \neq B_{21}$, whereas red lines represent the laser case $B_{12}=B_{21}$.
    Panel (b) depicts the relative population inversion from Eq.~\eqref{eq:population_inversion}, for the same two cases as in panel (a), with the dashed line highlighting equal population. In both panels the following parameter values were used for the purpose of illustration: $M=10^5$, $\Gamma_{\rm c} = 10^4\,\text{Hz}$, $\Gamma_{\rm nr} = 5\,\text{Hz}$ as well as $B_{12} = 0.5\,\text{Hz}$, $B_{21} = 1\,\text{Hz}$ for the photon BEC case and
    $B_{12} = 1\,\text{Hz} = B_{21}$ in the laser limit, respectively.}
	\label{fig:single_mode_model}
\end{figure}
is expected~\cite{Haken_book1985}.\\
As shown in Fig.~\ref{fig:single_mode_model}(a), the photon number for the single-mode photon Bose--Einstein condensate increases only weakly below a critical pumping value, while well above this threshold the photon mode occupation rises sharply. 
The dashed line represents the large-pumping limit of Eq.~\eqref{eq:Nminus_largep}. 
Specializing to the laser case requires that the emission and absorption rates are the same. 
Setting $B_{12}=B_{21}$ implies due to the KS relation~\eqref{eq:KS_relation} that the photon mode frequency coincides with the zero-phonon line, i.e.~$\omega_\ell = \omega_\text{ZPL}$. 
In this laser limit, as also shown in Fig.~\ref{fig:single_mode_model} (a), the photon number qualitatively behaves the same as in the photon BEC case. 
Note that in both cases the spontaneous emission is responsible for a finite photon occupation already below threshold~\cite{Haken_book1981,Haken_book1983,Haken_book1985,Risken_book1989}.\\
In order to distinguish both regimes we consider the population inversion $D=M_2 - M_1$~\cite{Haken_book1985}. 
Using Eqs.~\eqref{eq:molecule_solution1} and~\eqref{eq:molecule_solution2}, yields together with Eq.~\eqref{eq:photon_single_mode} the relative population inversion
\begin{equation}
	\label{eq:population_inversion}
	\frac{D}{M} = \frac{p- \Gamma_{\rm nr}- 2\Gamma_{\rm c} N_{+}/M}{\Gamma_{\rm nr} + p} \,,
\end{equation}
as shown in Fig.~\ref{fig:single_mode_model}(b). 
For small pumping strengths, the molecular population resides predominantly in the ground state. With increasing pumping power, the fraction of excited molecules grows. 
Well above some critical pumping value, the molecule populations saturate. 
The sign of the relative population inversion, i.e. whether it is positive or negative, allows a theoretical discrimination between the photon BEC and the laser regime, respectively. 
Evaluating Eqs.~\eqref{eq:population_inversion} by taking into account \eqref{eq:photon_single_mode} for large pumping yields 
\begin{equation}
	\label{eq:population_inversion-saturation}
\lim_{p \rightarrow \infty}	\frac{D}{M} = \frac{M(B_{12}- B_{21})+ 2\Gamma_{\rm c}}{M(B_{12}+B_{21})} \,.
\end{equation}
Thus, for the parameters chosen in Fig.~\ref{fig:single_mode_model}(b) for the photon BEC, where we have $B_{12} < B_{21}$, the majority of molecules are still in the ground state and, therefore, the saturation value for the relative population inversion~\eqref{eq:population_inversion-saturation} is negative. 
However, in case of the laser, where $B_{12} = B_{21}$ holds, we see from Eq.~\eqref{eq:population_inversion-saturation} that the majority of molecules are in the excited state. 
This saturated population inversion is characteristic for the lasing regime in agreement with Refs.~\cite{Haken_book1981,Haken_book1983,Haken_book1985}.
\section{Relevant Parameters}\label{sec3}
\begin{figure}[t]
	\centering
	\includegraphics[width=\columnwidth]{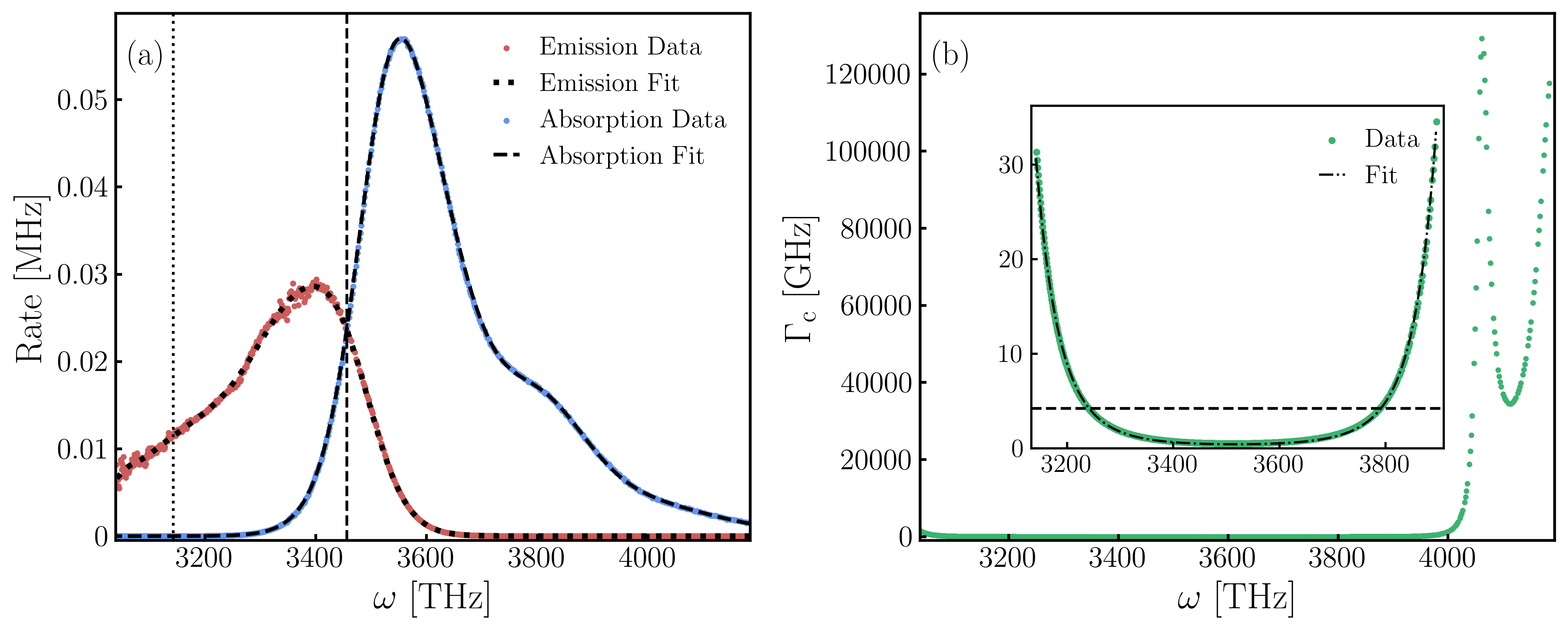}
	\caption{Experimental data reported in Ref.~\cite{Schmitt_Data2024}. Panel (a) displays the measured emission (red) and absorption (blue) rates $B_{21}(\omega)$ and $B_{12}(\omega)$ together with fits obtained from~\eqref{eq:spectra_fit_funca} and~\eqref{eq:spectra_fit_funcb}. The vertical dotted line indicates the cavity cut-off frequency $\omega_\text{cut}=2\pi\cdot 500\,\text{THz}$, while the dashed line marks the zero-phonon line $\omega_\text{ZPL}=2\pi\cdot 550\,\text{THz}$. Panel (b) shows the cavity loss rates. The inset magnifies the frequency region, which is most relevant for subsequent numerical simulations and includes the fit from Eq.~\eqref{eq:loss_fit}. The horizontal dashed line denotes the approximate value derived from Eq.~\eqref{eq:loss_approx}.}
	\label{fig:data_fit}
\end{figure}

The previous section introduced a thermalization model for a photon gas that depends on parameters, which have not yet been specified. 
Therefore, we now identify values for these parameters to enable subsequent numerical simulations. 
To this end, we follow the setup of Ref.~\cite{Weitz_Nature2010}, where a microcavity is filled with rhodamine 6G dye dissolved in ethylene glycol at room temperature. 
Typically, a total number of molecules of $M=10^8,\ldots,10^{10}$ is used in this experimental setup. 
Photons are effectively trapped in an isotropic harmonic confinement in two dimensions with frequencies $\omega_\ell = \omega_\text{cut} + l\cdot \Omega$ and degeneracies $g_\ell = \ell + 1$. 
Note that no polarization of the photons is taken into account. 
Here, the cut-off frequency $\omega_\text{cut}$ defines the lowest photon energy, and $\Omega$ denotes the energy-level spacing. 
Experimentally, condensation was observed for $\omega_\text{cut} \in \left[2\pi\cdot 500\,\text{THz},\ldots,2\pi\cdot526\,\text{THz}\right]$~\cite{Weitz_Nature2010,Sazhin_Natcom2024} and $\Omega = 2\pi\cdot 40\,\text{GHz}$~\cite{Weitz_Nature2010,Erglis_PRL2025}. 
Furthermore, non-radiative losses typically amount to $\Gamma_\text{nr} = 250\,\text{MHz}$~\cite{Sazhin_Natcom2024}.\\
Next, the emission and absorption spectra together with the cavity losses are considered. 
For the underlying setup, these parameters were measured in Ref.~\cite{Schmitt_Data2024}, satisfying the KS relation~\eqref{eq:KS_relation}, and are shown in Fig.~\ref{fig:data_fit}.
From these data, the zero-phonon line frequency is determined as the point where the two spectra coincide, as shown in Fig.~\ref{fig:data_fit}(a), yielding $\omega_\text{ZPL} = 2\pi\cdot 550\,\text{THz}$. 
For the numerical calculations it is necessary to know the explicit value of the spectra at certain frequencies. 
Therefore, we fit the experimental data. 
Inspired by Refs.~\cite{Radonjic_NJP2018,Kirton_PRL2013}, the main peak of the absorption spectrum was described by a Gaussian-like function. 
This function was derived from a Lindblad master equation under the assumption of having only a single vibrational state. 
We generalize this approach to contributions from multiple vibrational states~\cite{Kirton_PRA2015} using a superposition of $n=6$ Gaussians to fit the emission spectrum:
\begin{equation}
	\label{eq:spectra_fit_funca}
		B_{21}^{(\text{fit})}(\omega) = \sum_{i=1}^{n} A_i e^{-\frac{(\omega - \omega_i)^2}{2\sigma_i^2}}\,.
\end{equation}
Consistency with the KS relation~\eqref{eq:KS_relation} is ensured by constructing the corresponding absorption spectrum as the Kennard-Stepanov transform of the same Gaussian superposition:
\begin{equation}
	\label{eq:spectra_fit_funcb}
		B_{12}^{(\text{fit})}(\omega) = e^{\hbar\beta(\omega-\omega_\text{ZPL})}\cdot B_{21}^{(\text{fit})}(\omega)\,.
\end{equation}
From this point of view, the location of each Gaussian peak $\omega_i$ for $i=1, \ldots, n$ in Eq.~\eqref{eq:spectra_fit_funca} then corresponds to the frequency of a vibrational state, providing a natural generalization of the earlier approaches~\cite{Radonjic_NJP2018,Kirton_PRL2013}. \\
Figure~\ref{fig:data_fit}(b) seems to indicate that cavity losses are negligible over the relevant frequency range. 
However, the inset reveals frequency-dependent losses in the relevant interval, which are fitted with an inverse Gaussian function:
\begin{equation}
	\label{eq:loss_fit}
	\Gamma_\text{c}^{(\text{fit})}(\omega) = e^{a (\omega - b)^2 + c}\,.
\end{equation}
The cavity losses shown in Fig.~\ref{fig:data_fit}(b) reflect the characteristics of the specific cavity employed. 
To obtain an approximate description, which is applicable to a wider class of cavities, the profile is approximated by a frequency-independent value, following the approach of Refs.~\cite{Radonjic_NJP2018,Kirton_PRL2013,Kirton_PRA2015}.
This value is defined such that the areas under the two curves coincide, see Fig.~\ref{fig:data_fit}, yielding 
\begin{equation}
	\label{eq:loss_approx}
	\Gamma_\text{c}(\omega_\ell) \approx \Gamma_\text{c} = \frac{1}{\omega_\text{max} - \omega_\text{cut}} \int_{\omega_\text{cut}}^{\omega_\text{max}} d\omega\, \Gamma_\text{c}^\text{(fit)}(\omega)\,,
\end{equation}
where $\omega_\text{max}=3800\,\text{THz}$ is the maximal frequency considered. 
This allows the cavity loss profile to be approximated by $\Gamma_\text{c} = 3.5\,\text{GHz}$.
\section{Numerical Simulation and Discussion}\label{sec4}
Section~\ref{subsec22} showed that the system of rate equations~\eqref{eq:photon_rate}--\eqref{eq:molecule_cons} admits in the steady state a formal solution. 
However, quantitative predictions for observables such as the chemical potential and its dependence on the system parameters require solving the complete set of equations~\eqref{eq:open_BEC} and~\eqref{eq:molecule_frac}.
To this end, this section presents numerical simulations addressing three representative cases.\\
The first case examines the effect of the number of thermal photon modes $\ell_\text{total}$ included in the simulation. 
Increasing the number of modes demonstrates convergence toward the thermodynamic limit and agreement with the experiment.
The second case investigates the role of the cavity. 
Varying the cavity loss rate $\Gamma_{\rm c}$ reveals how the chemical potential and the critical particle number for achieving condensation change in response.
The third case analyzes the influence of the dye solution by modifying the non-radiative loss parameter $\Gamma_\text{nr}$. 
In all three scenarios, the system of rate equations is solved using a Newton algorithm, with all other parameters fixed to isolate the effect of the parameter under study.
Moreover, in the following the cavity cut-off frequency is set to $\omega_\text{cut}=2\pi\cdot 500\text{ GHz}$ and a total molecule number of $M=10^{10}$ is considered.   
\subsection{Thermodynamic Limit}\label{subsec41}
First, the effect of the total number of thermal photon modes $\ell_\text{total}$, which are taken numerically into account, is analyzed. 
This represents the basis for the subsequent discussion, as it determines the dimensionality required for the calculations. 
To assess the influence of $\ell_\text{total}$, the cavity loss rate is approximated by $\Gamma_\text{c} = 3.5\,\text{GHz}$, as described at the end of Sec.~\ref{sec3}. 
In addition, the non-radiative loss rate is set to $\Gamma_\text{nr} = 0\,\text{MHz}$, as it is not expected to influence significantly the occupation of the photon modes. 
A detailed justification for this assumption is provided below in Sec.~\ref{subsec43}. 
Furthermore, in contrast to the experiment, the pumping strength $p$ is assumed to be known in this numerical analysis, as it is the control parameter of the system, which is increased from small to large values.
As a consequence, this leads to an increase of the total photon number
\begin{equation}
	\label{eq:N_toatal}
	N_\text{total} = \sum_{\ell=0}^{\ell_\text{total}} N_\ell\,.
\end{equation}
Thus, the dependence of any observable on the pumping strength $p$ can be converted into a function of the total photon number~\eqref{eq:N_toatal}. 
This conversion makes it possible to directly compare simulation results with experimental measurements.\\
Solving in steady state 
\begin{figure}[t]
	\centering
	\includegraphics[width=\columnwidth]{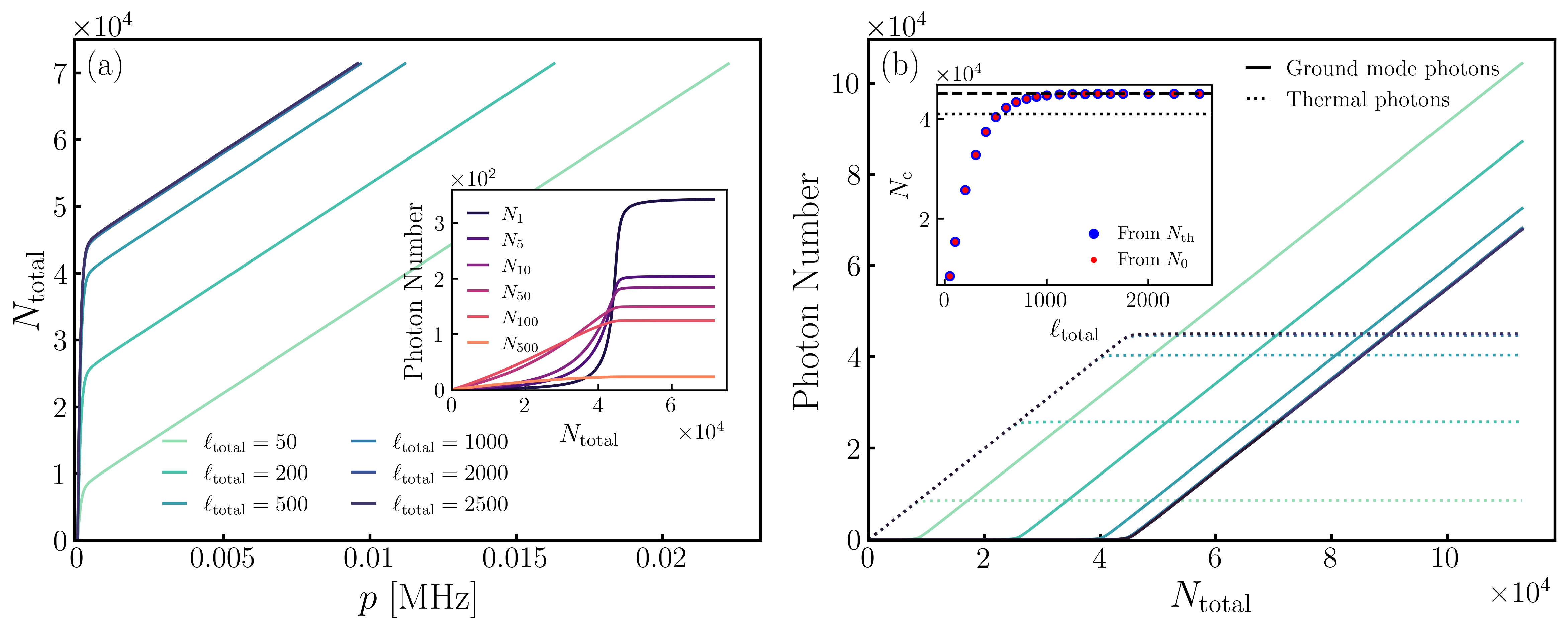}
	\caption{Panel (a) shows the total photon number as a function of the pumping strength. The inset displays the occupation for selected modes. Panel (b) represents the occupations of the ground state and of the thermal states. The inset shows the critical particle number obtained independently from ground-state (red) and thermal-state (blue) occupations. The dashed horizontal black line indicates the numerically calculated thermodynamic-limit value, whereas the dotted horizontal black line depicts the thermodynamic-limit value of the closed system calculated from Eq.~\eqref{eq:Nc_closed}. The color coding is identical in both panels.}
	\label{fig:photons_change_modes}
\end{figure}
the coupled algebraic equations ~\eqref{eq:open_BEC} and~\eqref{eq:molecule_frac} yields the resulting total photon number~\eqref{eq:N_toatal} as shown in Fig.~\ref{fig:photons_change_modes}(a). 
It increases steeply at small pumping strengths until the critical particle number $N_{\rm c}$ is reached, after which it grows linearly. 
Notably, the linear slope is the same for different values of the total photon mode number. 
Furthermore, the critical particle number $N_{\rm c}$, and thus photon condensation, is reached already at relatively small pumping strengths. 
Figure~\ref{fig:photons_change_modes}(b) shows the photon occupation of both ground state $N_0$ and thermal state
\begin{equation}
	\label{eq:thermal_photons}
	N_\text{th} = \sum_{\ell=1}^{\ell_\text{total}} N_\ell\,.
\end{equation}
For small photon numbers, only thermal states are occupied. 
Once the critical particle number $N_{\rm c}$ is reached, the thermal states saturate, as shown in the inset of Fig.~\ref{fig:photons_change_modes}(a), and the ground state becomes occupied. 
Thus, the open-dissipative Bose--Einstein distribution~\eqref{eq:open_BEC} yields the same qualitative result for the occupations of ground and thermal state as for the standard Bose--Einstein distribution of a closed system. 
In this case, while neglecting the impact of spontaneous emission, the occupations are empirically approximated according to \cite{Pethick_Smith,Pitaevskii_Stringari}
\begin{eqnarray}
	\label{eq:photon_occupationa}
		N_0(N_\text{total}) &=& \Theta(N_\text{total} - N_\text{c})\Big[N_\text{total} - N_\text{c}\Big]\,,\\
        \label{eq:photon_occupationb}
		N_\text{th}(N_\text{total}) &=& N_\text{total} - N_0(N_\text{total})\,,
\end{eqnarray}
where $\Theta$ denotes the Heaviside function. 
Thus, there are two complementary approaches for defining and calculating the critical particle number $N_{\rm c}$. 
The first one relies on the saturation value of the thermal states, whereas the second one is based on determining the particle number at which the ground state becomes occupied. 
In the latter approach, however, one must account for the fact that~\eqref{eq:photon_occupationa} neglects the impact of the spontaneous emission, so that the ground state occupation does not exhibit a sharp onset, as shown Fig.~\ref{fig:photons_change_modes}(b). 
Instead, spontaneous emission leads to partial ground-state occupation already below $N_\text{c}$. 
To address this issue, the large-photon-number behavior of the ground state is linearly fitted, and $N_\text{c}$ is extracted from this fit by determining the zero crossing. 
The results of both approaches are shown in the inset of Fig.~\ref{fig:photons_change_modes}(b), which turn out to be consistent. 
The inset further demonstrates that the critical particle number $N_{\rm c}$ increases with the inclusion of additional photon modes, but saturates for $\ell_\text{total}\approx 1500$. 
This saturation indicates that the thermodynamic limit is reached, yielding the critical particle number $N_\text{c}\approx 45052$. 
This value lies well within the error bars of the original experiment~\cite{Weitz_Nature2010}. 
In addition, this critical particle number can also be compared with the closed-system case, where all open-dissipative parameters vanish. 
Considering the thermodynamic limit together with the first finite-size correction, one obtains for a two-dimensional Bose gas confined in an isotropic harmonic trap with frequency $\Omega$ the expression~\cite{Klunder_EPJ2009}
\begin{equation}
	\label{eq:Nc_closed}
	N_\text{c}^{\text{(closed)}} = \left(\frac{1}{\hbar\beta\Omega}\right)^2\zeta(2)\left[1 + \frac{\gamma - \frac{1}{2} - \ln(\hbar\beta\Omega)}{\frac{\zeta(2)}{\hbar\beta\Omega} - \frac{1}{2}}\right]\,.
\end{equation}
Here, the Riemann zeta-function has the value $\zeta (2)=\pi^2/6$ and $\gamma=0.5772\ldots$ denotes the Euler-Mascheroni constant. 
By inserting the trap frequency $\Omega = 2\pi\cdot 40\,\text{GHz}$, the critical particle number is obtained at room temperature to be $N_\text{c}^{\text{(closed)}}\approx 40976$. 
This finding for the standard Bose--Einstein distribution of a closed system differs significantly from the critical particle number $N_\text{c}\approx 45052$ determined from the open-dissipative Bose--Einstein distribution. 
Thus, this comparison provides clear evidence that the additional term in~\eqref{eq:open_BEC} is important 
\begin{figure}[t]
	\centering
	\includegraphics[width=\columnwidth]{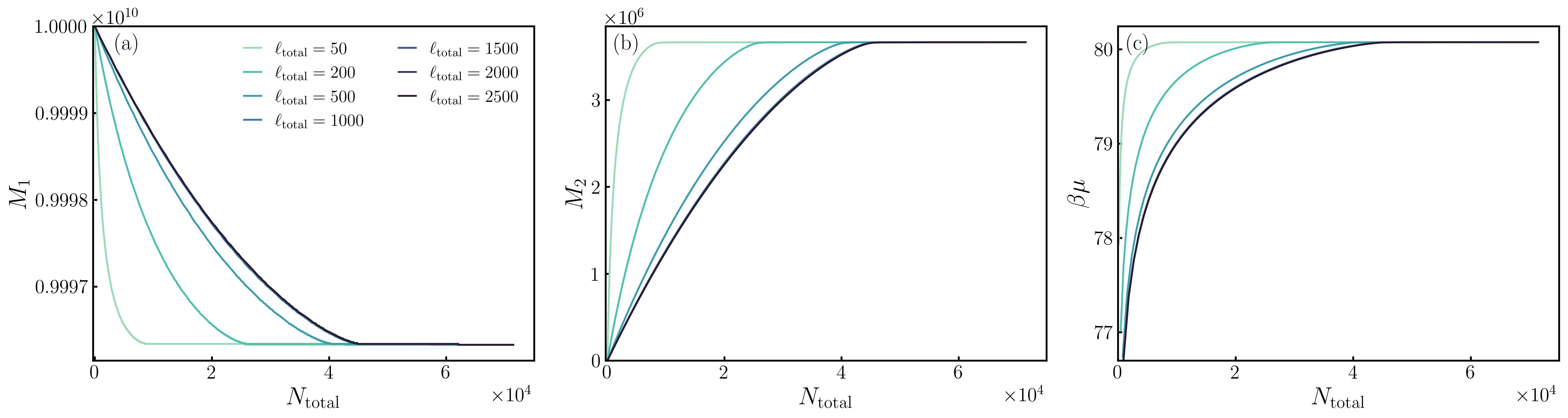}
	\caption{Panel (a) and (b) show the occupation of the ground state and excited state molecules, whereas panel (c) depicts the chemical potential, calculated via Eq.~\eqref{eq:chemical_pot}.}
	\label{fig:chem_pot_change_modes}
\end{figure}
and should not be neglected. However, this presents a paradox that must be resolved. In experiments, the additional term is typically neglected, and results are compared directly to the standard Bose--Einstein distribution~\cite{Klaers_Nature2010,Klaers_APB2011,Klaers_PRL2012,Schmitt_PRA2015,Damm2016,Busley_PRA2023}. These comparisons show good agreement, suggesting, in contrast to our claim, that the additional term has a negligible influence. This paradox can be resolved by considering the experimental measurement uncertainty, which is currently on the order of $10\%$~\cite{Redmann_PRL2024,Busley_Science2022}. This level of uncertainty is large enough to obscure the subtle effects predicted by the open-dissipative Bose--Einstein distribution.\\
Although the open-dissipative nature of the system strongly affects the critical particle number, it is necessary to verify whether the same holds also for the chemical potential. 
For this purpose, ground- and excited-state molecule populations are considered first. 
As shown in Fig.~\ref{fig:chem_pot_change_modes}(a), the ground-state population decreases with increasing photon number and then saturates at the critical particle number. 
Due to the conservation of molecules in Eq.~\eqref{eq:molecule_cons}, the excited-state population increases correspondingly until it also saturates; see  Fig.~\ref{fig:chem_pot_change_modes}(b). 
Note that it is evident that here no laser behavior is achieved. 
Since the ground-state population is significantly larger than the excited-state population for all pump parameters, the relative population inversion~\eqref{eq:population_inversion} remains negative. According to Sec.~\ref{subsec23}, therefore, the lasing regime never occurs. 
The resulting chemical potential~\eqref{eq:chemical_pot}, shown in Fig.~\ref{fig:chem_pot_change_modes}(c), increases with increasing photon number and saturates once the critical photon number is reached. 
In the condensed phase, the saturated chemical potential follows
\begin{equation}
	\label{eq:chem_pot_cond}
	\beta \mu_{\rm c} = \hbar\beta\omega_\text{cut} - \ln\left[1 - \frac{\Gamma_\text{c}}{B_{21}(\omega_\text{cut}) M_2}\right]\,,
\end{equation}
which yields $\beta \mu_{\rm c}\approx 80.08$, corresponding to the numerically found value in Fig.~\ref{fig:chem_pot_change_modes}(c). 
Nevertheless, the chemical potential still reflects the influence of the open-dissipative nature of the system. 
Taking the closed system limit yields $\lim\limits_{\Gamma_\text{c} \rightarrow 0}\beta \mu^{(\text{closed})}_{\rm c} = \hbar\beta\omega_\text{cut} \approx 79.98$. 
Thus, the chemical potential turns out to be only marginally influenced by the open-dissipative property of the system.
\subsection{Cavity Influence}\label{subsec42}
In the previous subsection it was shown that the open-dissipative character of a photon BEC has a pronounced effect on the critical particle number $N_\text{c}$, while its influence on the critical chemical potential $\mu_{\rm c}$ is small. 
To explore this contrast in more detail, the present section studies how both $N_\text{c}$ and $\mu_{\rm c}$ vary with cavity losses $\Gamma_{\rm c}$. 
To isolate this dependence, the non-radiative losses are neglected again. Moreover, the number of thermal photon modes is chosen to be $\ell_\text{total}=2000$, ensuring that the thermodynamic limit is reached, as established in the previous subsection. 
\begin{figure}[t]
	\centering
	\includegraphics[width=\columnwidth]{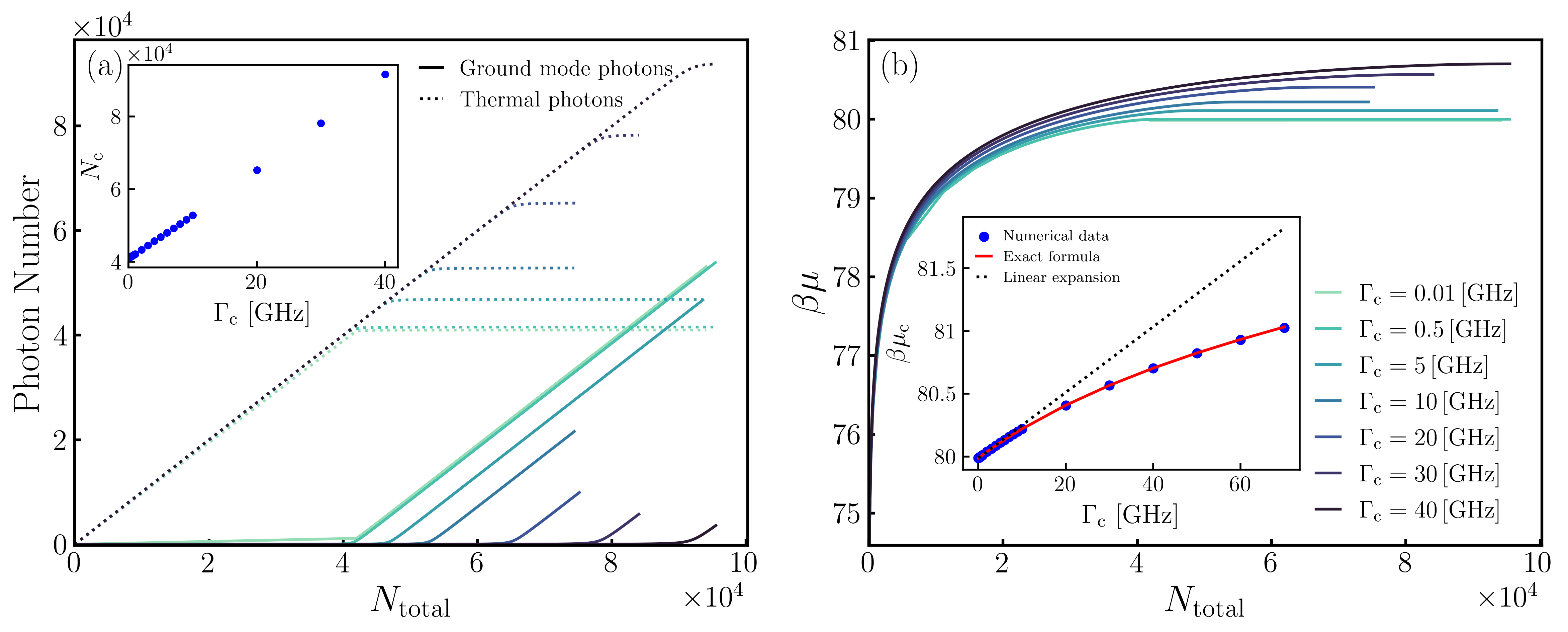}
	\caption{Panel (a) depicts the photon occupation of ground state and thermal states for varying cavity losses $\Gamma_{\rm c}$. The inset illustrates the corresponding dependence of the critical particle number. Panel (b) presents the chemical potential. Its inset highlights the value in the condensed phase, where dots indicate numerical results, the red curve depicts the exact expression of Eq.~\eqref{eq:chem_pot_cond}, and the dotted black line shows the expansion of Eq.~\eqref{eq:chem_pot_cond}
	valid for small cavity losses according to Eq.~\eqref{eq:mu_cond_expansion}.}
	\label{fig:change_losses}
\end{figure}
The solution of the  steady-state rate equations~\eqref{eq:open_BEC} and~\eqref{eq:molecule_frac} for varying cavity losses $\Gamma_{\rm c}$, shown in Fig.~\ref{fig:change_losses}(a), reveals clear trends. 
As cavity losses $\Gamma_{\rm c}$ increase, the number of condensed, i.e.~coherent, photons decreases, while the thermal,  i.e.~incoherent, occupation increases. 
Consequently, the critical particle number $N_{\rm c}$ increases linearly with the cavity losses $\Gamma_{\rm c}$, as illustrated in the inset of Fig.~\ref{fig:change_losses}(a). 
In particular, a tenfold change in $\Gamma_\text{c}$ results in a change of a similar order of magnitude for $N_\text{c}$. 
This confirms that the equilibrium BEC distribution is insufficient for photon condensates, which exist in non-equilibrium and must therefore  be described by the open-dissipative BEC distribution of Eq.~\eqref{eq:open_BEC}.
In the limit $\Gamma_\text{c} \rightarrow 0$, the value $N_\text{c} \approx 40985$ is obtained, which is in excellent agreement with the prediction of the closed system from Eq.~\eqref{eq:Nc_closed}. 
It should be noted, that for the smallest loss rate in Fig.~\ref{fig:change_losses}(a), the ground state appears to be slightly occupied even far below the threshold. This apparent occupation, however, can be attributed to the numerical precision.\\
The chemical potential, shown in Fig.~\ref{fig:change_losses}(b), exhibits a much weaker dependence on cavity losses. 
However, although its variation is small, the inset highlights that in the condensed phase, $\mu$ follows the logarithmic dependence predicted by Eq.~\eqref{eq:chem_pot_cond}, in contrast to the linear scaling of $N_\text{c}$. 
Moreover, the behavior diverges markedly from the linear approximation for small cavity losses:
\begin{equation}
	\label{eq:mu_cond_expansion}
	\beta\mu_{\rm c} = \hbar\beta\omega_\text{cut} + \frac{\Gamma_\text{c}}{B_{21}(\omega_\text{cut}) M_2(\Gamma_\text{c} = 0)} + \mathcal{O}(\Gamma_\text{c}^2)\,.
\end{equation}
Taken together, these results indicate that the appropriate statistical description of photon condensation is provided by the open-dissipative BEC distribution~\eqref{eq:open_BEC}, rather than by the standard Bose--Einstein distribution. 
\subsection{Influence of Non-radiative Losses}\label{subsec43}
Finally, the analysis focuses on the impact of the non-radiative loss rate $\Gamma_{\rm nr}$. 
To isolate this effect, the cavity loss rate is fixed to $\Gamma_\text{c} = 3.5\,\text{GHz}$, while $\ell_\text{total} = 2000$ photon modes are included to guarantee that the thermodynamic limit is reached, as established in Sec.~\ref{subsec41}.
\begin{figure}[t]
	\centering
	\includegraphics[width=\columnwidth]{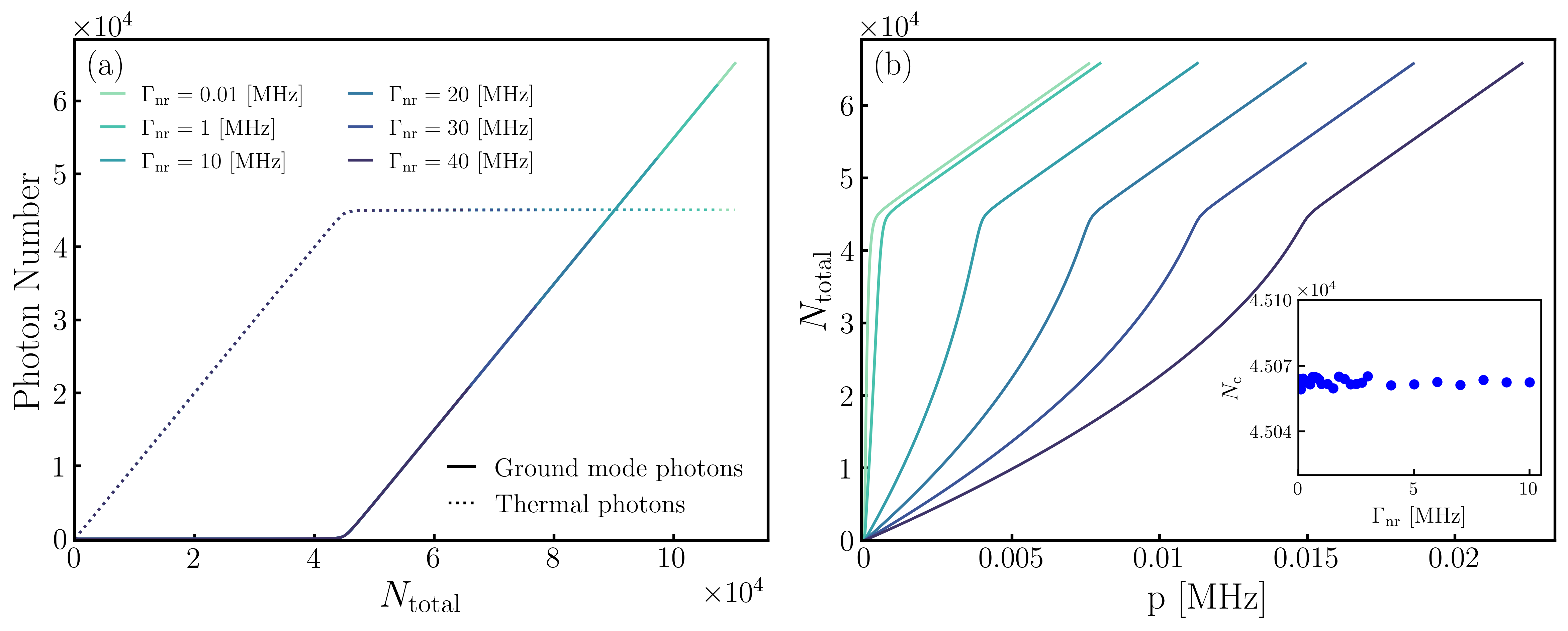}
	\caption{Panel (a) shows the occupation of ground-state and thermal-state photons for different non-radiative loss rates. Panel (b) presents the dependence of the total photon number on the pumping strength for the same loss rates. The inset highlights that the critical particle number does not change with varying non-radiative losses.}
	\label{fig:change_nonrad_losses}
\end{figure}
Figure~\ref{fig:change_nonrad_losses}(a) demonstrates that the non-radiative losses do not alter the occupations of ground-state or thermal-state photons. 
As a consequence, both the molecular occupations and the chemical potential remain unchanged. 
The influence of the radiationless loss rate $\Gamma_{\rm nr}$ appears instead in the total photon number $N_{\rm total}$, as is shown in Fig.~\ref{fig:change_nonrad_losses}(b). 
With increasing non-radiative loss rate $\Gamma_{\rm nr}$, the pumping strength $p$ required to achieve condensation increases, since a larger fraction of molecules is lost and cannot contribute to condensation. 
In contrast, the critical particle number number $N_{\rm c}$ is not affected. 
Note that the minimal changes in $N_{\rm c}$, as shown in the inset of Fig.~\ref{fig:change_nonrad_losses}(b) can be attributed to the numerical precision.
All these findings justify a posteriori having neglected non-radiative losses in the previous two subsections, as the dependence of all examined quantities on the total photon number is not affected.
\section{Summary and Outlook}\label{sec5}
This work introduced a rate-equation framework for describing the thermalization of photon Bose--Einstein condensates. 
Solving the model in steady state yields the open-dissipative Bose--Einstein distribution, which provides the appropriate statistical description of photon condensation. 
This distribution extends the standard Bose--Einstein form by including an additional term that accounts for the open character of the underlying system. 
Numerical simulations with experimentally realistic parameters show that the thermodynamic limit is reached once a sufficiently large number of photon modes are included. 
Moreover, a variation of the cavity loss rate reveals a larger and a smaller effect on the critical particle number and the chemical potential, respectively. 
These findings establish that photon Bose--Einstein condensates are correctly described by the open-dissipative rather than by the standard Bose--Einstein distribution.\\
It would be worth measuring this difference between the open-dissipative and the standard Bose--Einstein distribution in an experimental set-up. 
As the difference in the critical photon number $N_{\rm c}$ can amount to 10~\% or more for experimentally realistic parameters, this seems to be a realistic goal.
\section*{Data availability statement}
The data supporting the findings of this study are available upon reasonable request from the authors.

\section*{Acknowledgments}
We thank Fl{\'a}via Braga Ramos, Nikolai Kaschewski, Milan Radonji{\'c}, and Julian Schmitt for insightful discussions. Furthermore, financial support by the Deutsche Forschungsgemeinschaft (DFG, German Research Foundation) is acknowledged via the Collaborative Research Center SFB/TR185 (Project No. 277625399).

\bibliography{references.bib}

\end{document}